\documentclass[pra,aps,twocolumn]{revtex4}
\usepackage{graphicx} 
\input psfig.sty 
 
\newcommand{\la}{\langle} 
\newcommand{\ra}{\rangle} 
\newcommand{\ba}{\begin{eqnarray}} 
\newcommand{\ea}{\end{eqnarray}} 
\newcommand{\be}{\begin{equation}} 
\newcommand{\ee}{\end{equation}}

\begin{document} 
 
\title{Quantum Error Correction of a Qubit Loss in an Addressable Atomic System} 
 
\author{J. Vala$^{1}$, K. B. Whaley$^{1}$ and D. S. Weiss$^{2}$} 
 
\affiliation{ 
$^1$ Department of Chemistry and Pitzer Center for Theoretical Chemistry, University of California, Berkeley, California 94720\\ 
$^2$ Department of Physics, Pennsylvania State University, University Park, Pennsylvania 16802-6300 
} 
 
\date{\today} 
 
\begin{abstract} 
We present a scheme for correcting qubit loss error while 
quantum computing with neutral atoms in an addressable optical lattice. The  
qubit loss is first detected using a quantum non-demolition measurement and  
then transformed into a standard qubit error 
by inserting a new atom in the vacated lattice site. The logical qubit,  
encoded here into four physical qubits with the Grassl-Beth-Pellizzari 
code, is reconstructed via a sequence of one projective measurement,  
two single-qubit gates, and three controlled-NOT operations. No ancillary 
qubits are required. Both quantum 
non-demolition and projective measurements are implemented using a cavity 
QED system which can also detect a general leakage error and thus allow 
qubit loss to be corrected within the same framework.
The scheme can also be applied in quantum computation with trapped ions or with photons. 
\end{abstract} 
 
\maketitle 
 
Isolated neutral atoms in their electronic ground states are ideal qubits. Atomic coherences can live for much longer than observation times, and atoms in traps can be observed for many seconds. Trap perturbations on atomic coherences, as well as other interactions with their environment, can be minimal and can be characterized extremely well. Single qubit gates can be nearly perfectly executed on neutral atoms, and there are several promising approaches to mutually entangling pairs or groups of neutral atoms \cite{Jaksch:99,Brennen:99,Jaksch:00}. Large numbers of neutral atoms can be arrayed in optical 
lattices or other atom traps \cite{Vala:03,Sebby-Strabley:05}. In short, neutral atoms can demonstrably 
meet all the criteria for scalable quantum computation \cite{DiVincenzo:00}.  For reliable computation, one requires also the ability to correct quantum errors.  Here an unusual situation arises for neutral atoms trapped in optical lattices because background gas collisions can eject the atoms, causing the qubit to simply disappear from the system. A cryogenic environment can drastically minimize background gas collisions \cite{Libbrecht:95}, but realizing reliable quantum computation in optical lattices will require correction of such qubit loss errors.

In this paper we present a scheme to correct qubit loss error during quantum computation with neutral atoms. We specifically consider neutral atoms in an addressable optical lattice, although the method could be applied to any neutral atom system, or in fact, to any quantum computing system that might experience qubit loss. The key idea is to translate these non-standard quantum errors into a standard quantum error model, and then to correct them with known quantum error correction schemes (e.g. those based on the Calderbank-Shor-Steane (CSS) code \cite{Calderbank:96,Steane:96b} or the Grassl-Beth-Pellizari (GBP) code for erasure error \cite{Grassl:97}). The error is detected by a quantum non-demolition (QND) measurement, implemented using a cavity QED system \cite{Mabuchi:99} with high numerical aperture (N.A.) optics so that the cross section of the mode at its center is much less than the cross sectional area of a lattice cell. The measurement must identify whether an atom is present at the lattice site without resolving the qubit quantum levels. Once the error is identified, the vacated lattice site is filled with a new atom in the qubit ground state using an optical tweezer~\cite{Grier:03}. At this point the net quantum error is equivalent to a standard qubit error due to spontaneous emission of a photon, so it can be corrected by standard quantum error correction. In our specific example, the logical qubit, encoded into four physical qubits of the GBP code, is reconstructed via a sequence of one projective measurement, two single-qubit gates and three controlled-NOT operations. A significant feature of the present scheme is that no ancillary qubits are required. 

\paragraph*{System.}
The quantum computer under consideration consists of a lattice of neutral atoms, here $^{87}Rb$, trapped in perpendicular standing waves of linearly polarized laser beams \cite{Vala:03,Weiss:03}. The lattice is characterized by a large lattice constant $a = 5 \mu m$ and is therefore addressable, meaning that each atom can be individually controlled by an optical field. Because of this requirement, the analysis is simplest for a two-dimensional lattice. 
It is initialized into a perfect lattice with a single atom per lattice site, each in a given internal state \cite{Vala:03,Weiss:03}. 

The qubit can be any two magnetic hyperfine sublevels $m_F$ of the electronic $5^2S_{1/2}$ ground state hyperfine manifolds characterized by the total angular momentum of the atom  $F = 1$ and $F = 2$. We elect to use the field insensitive magnetic hyperfine states, $m_F = 0$,  
\ba 
|0\ra & = & |5^2S_{1/2}, F = 1, m_F = 0\ra \nonumber \\ 
|1\ra & = & |5^2S_{1/2}, F = 2, m_F = 0\ra. 
\ea 
This choice of qubit states strongly suppresses qubit dephasing due to fluctuations of lattice or magnetic fields.

\paragraph*{Errors.}
A quantum computer based on neutral atoms trapped in an addressable optical lattice is exposed to two types of errors: (i) qubit errors, which may be accounted for by standard quantum error correction, and (ii) general leakage errors. A quantum leakage error is an uncontrolled rotation of the qubit state vector $|\psi\ra$ from the qubit subspace ${\mathcal{H}}_{\psi}$ to the rest of the physical Hilbert space ${\mathcal{H}}_{\psi}^{\bot}$. There are two types of leakage: non-qubit atomic hyperfine levels can be populated for example by spontaneous emission or imperfect quantum computing operations; and atoms can be lost from the lattice due to collisions with background gas particles. Both processes typically occur on a timescale from $\sim$ 10 to 100 s. 

\paragraph*{Error Correction.}
The essential first step in correcting a qubit loss is to transform the loss into a standard quantum error using a sequence of physical operations. The standard errors can then be corrected with a known quantum error correction scheme. Formally, the process of qubit loss and its correction can be described as the following sequence: 
\ba \label{Eq:Sequence} 
(1)&~\rho_{\psi} \otimes \rho_{00,out}^{\otimes n} &\to \nonumber \\  
(2)&\rho_{err,i} \otimes \rho_{00,out}^{\otimes n} &\to \nonumber \\  
(3)&\rho_{err,i} \otimes \rho_{00,i} \otimes \rho_{00,out}^{\otimes (n-1)} &\to \nonumber \\  
(4)&\rho_{\psi} \otimes \rho_{00,out}^{\otimes (n-1)}& 
\ea 
Here $\rho_{\psi} = |\psi\ra\la\psi|$ is the density matrix describing the pure state of the quantum computer, and $\rho_{err,i}$ is the quantum computer state after the qubit loss error at the site $i$ is detected.
$\rho_{00,out}$ is the ground state atom prepared outside the computational lattice, and 
$\rho_{00,i}$ describes this atom when inserted onto the vacated lattice site $i$.  
Stage (1) describes the initial pure state of the quantum computer that, in stage (2), loses a qubit at the site  
$i$. The QND measurement results in the ejected atom being 
traced out from the density matrix, so that after stage (2) 
$\rho_{err,i} = Tr_i \rho_{\psi}$.   QND measurements are made periodically 
on all atoms in order to identify the presence or 
absence of atoms at each site.  These measurements leave the qubit  
states unperturbed. At stage (3), a  
new atom in the qubit ground state is inserted at the empty site by action 
of a source that is conditional on the QND measurement. The effect of this  
sequence is to transform the qubit loss to an error that is equivalent 
to spontaneous emission (also known as amplitude damping,  
see e.g. \cite{Nielsen:00}). This error is then corrected in stage (4) by  
standard quantum error correction methods. 
 
\begin{figure} 
\includegraphics[width=3in] {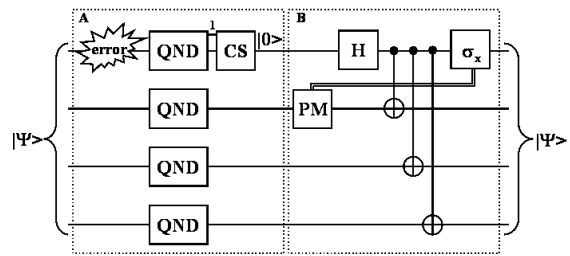} 
\caption{Qubit loss error correction circuit. In this example, 
the first qubit  
of the GBP code is lost to the environment. Quantum non-demolition  
measurements (QND) indicate the location of the error and trigger the  
conditional source (CS) to inject a new atom in state $|0\ra$ into the optical  
lattice site. A projective measurement (PM) purifies the mixed state resulting  
from tracing out the lost qubit. Subsequent single qubit and two qubit  
operations reconstruct the encoded logical qubit.} 
\label{Fig1} 
\end{figure} 
 
The quantum error described above is equivalent to an erasure error because its location is revealed by the QND measurement. We seek an encoding that satisfies two criteria: (i) all states of code words stay distinguishable when one physical qubit is lost, and (ii) the code words then preserve coherences of the encoded logical qubit.
A quantum erasure encoding satisfying these requirements was previously  
developed in \cite{Grassl:97} and applied to error correction for photon loss  
in \cite{Gingrich:03} in a scheme that encoded two logical qubits and used 
two ancillas. Here we also use the GBP code of \cite{Grassl:97} to encode one  
logical qubit and show that this can be combined with the above error
correction procedure to correct loss of single qubits without ancillas. 
The correction procedure proposed here is specially tailored for experimental 
implementation with neutral atoms in an addressable optical lattice; 
it avoids ancillary qubits and uses only a small number of operations.

The logical qubit $|\psi\ra_L = c_0 |0\ra_L + c_1 |1\ra_1$ is encoded as 
\ba 
|0\ra_L & = & \frac{1}{\sqrt{2}} (|0000\ra + |1111\ra) \nonumber \\ 
|1\ra_L & = & \frac{1}{\sqrt{2}} (|0011\ra + |1100\ra). 
\ea 
Using the density matrix notation of \cite{Gingrich:03}, i.e. 
$\rho = \Sigma_k^m p_k |\phi_k><\phi_k| = \{(\phi_1,p_1) ... (\phi_m,p_m)\}$, 
the initial state of the logical qubit is given by  
\be 
\rho_i = \{(\frac{1}{\sqrt{2}} [c_0 (|0000\ra + |1111\ra) + c_1 (|0011\ra + |1100\ra)], 1)\}. 
\label{Eq:Initial}
\ee 
 
We now demonstrate the error correction procedure in detail for the case when 
qubit loss occurs in the first physical qubit position. Losing the first 
qubit into the environment causes the density matrix to be traced over this  
qubit and results in an equal mixture of two pure state projections 
onto the  
even and odd bit-string parity sector of the Hilbert space of the remaining  
three qubits. After inserting a new ground state qubit atom 
into the vacant site with 
the conditional source, the mixed state then becomes 
\be 
\rho_e = \{(c_0 |0000\ra + c_1 |0011\ra, \frac{1}{2}),(c_0|0111\ra + c_1 |0100\ra, \frac{1}{2})\}. 
\label{Eq:Mixed} 
\ee 
 
We note that the second physical qubit provides a label for the two pure state projections in (\ref{Eq:Mixed}). The mixed state can then be purified by projective measurement of this qubit. Thus, if the projective measurement result on the second qubit is 0, the state  
$\rho_e$ is collapsed into the pure state  
$\rho_p = \{(c_0 |0000\ra + c_1 |0011\ra, 1)\}$. Applying a Hadamard  
gate on the first qubit gives the superposition
\be
\{\frac{c_0}{\sqrt{2}}(|0000\ra + |1000\ra) + \frac{c_1}{\sqrt{2}}(|0011\ra + |1011\ra),1\}.
\label{Eq:Hadamard}
\ee
This state is then transformed back into the initial state $\rho_i$ (Eq. (\ref{Eq:Initial})) by CNOT gates which are conditional on the first qubit and act on each of the other qubits (i.e. if the first qubit is 1, then all other qubits are flipped). 
If however the projective measurement on the second qubit yields 1, 
the state $\rho_e$ is collapsed onto  
the pure state $\rho_p = \{(c_0 |1111\ra + c_1 |0100\ra, 1)\}$. The same  
sequence of operations as in the previous case, i.e. the same Hadamard and CNOT gates, results in the state
\be
\{\frac{c_0}{\sqrt{2}}(|1000\ra + |0111\ra) + \frac{c_1}{\sqrt{2}}(|1011\ra + |0100\ra),1\}.
\ee
which becomes the initial state $\rho_i$ after the first qubit is flipped by $\sigma_x$ operation conditional on result of the projective measurement on the second qubit. In both cases, the same sequence of operations, illustrated by error correction circuit shown in Fig. \ref{Fig1}, leads to the complete reconstruction of the initial state and thus correction of the quantum error due to qubit loss.
 
This qubit loss error correction scheme can also accomodate a leakage  
error. The QND measurement in principle allows one to identify that quantum  
states have leaked out of the qubit Hilbert space, without perturbing the  
qubit states. It can hence act as a projective (von Neumann) measurement that  
collapses a leaked state into a particular detected state that can then be  
tranformed back to the qubit ground state via a sequence of unitary 
operations. 
The leakage error is thereby transformed to a qubit error that can be dealt  
with by standard quantum error correction, just as was done for qubit loss  
above.

We now discuss physical implementation of the various elements required for  
this qubit loss error correction scheme, namely QND measurements, projective
measurements, and a conditional source. 
 
\paragraph*{Error Detection.} 
 
\begin{figure} 
\includegraphics[width=3in] {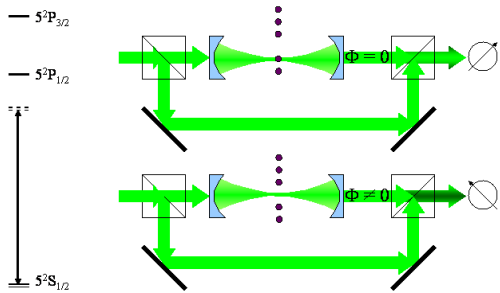} 
\caption{Scheme illustrating the implementation of the quantum non-demolition measurement using the cavity QED system.} 
\label{Fig2} 
\end{figure}

Identification of qubit loss error is the central component of the scheme.  
This can be achieved with a QND measurement of the presence or absence of  
atoms on individual optical lattice sites using a cavity QED system \cite{Mabuchi:99} with a small mode volume. 
In particular, when the optical field is far red-detuned from the excited state 
$5^2P_{1/2}$, the measurement outcome distinguishes only whether an atom is  
present in the field. It does not resolve the qubit levels, identified with  
the field-insensitive magnetic hyperfine  $5^2S_{1/2}$ states ($m_F = 0$),  
which impose nearly identical distinct phases to the cavity optical field. 
The measurement is illustrated in Fig \ref{Fig2}.
We remark that the measurement imposes a small deterministic relative phase  
onto the qubit levels which can be inverted by an additional 
single qubit operation or simply taken into account in the quantum compiler.

The observable of this QND measurement  
is the phase shift exerted on the cavity field due to the 
interaction with an atom. 
The total phase shift, $\phi$, is determined by the finesse of the cavity, $f$: 
$\phi = f \phi_1$ where $\phi_1$ is the phase shift for a single 
scattering event. 
$\phi_1$ is given by  
$\phi_1 = \frac{D_0}{4 \delta/\Gamma}$, 
with $\Gamma$ the spontaneous emission rate, $\delta$ the detuning 
of the field from the transition frequency, and 
$D_0$ the resonant optical density, which is proportional 
to the resonant cross-section $\sigma_0 = \frac{\lambda^2}{2\pi}$.  
$D_0$ is equal to $n L \sigma_0$ \cite{Loudon:83}, where  
$n L = \frac{C}{\sigma}$, with the geometric factor $C$ 
defined for tight focus as an overlap of the Gaussian beam 
of width $w$ in the plane perpendicular to the direction of  
propagation,\ with the cylinder given by the atomic cross section $\sigma$
of the radius $w_0$: $C = \frac{\pi}{4} erf(\frac{w/w_0}{\sqrt{2}})$. 
The calculation of the phase shift may include additional geometric factors
arising from the atomic dipole moment orientation. High N.A. cavity optics
can be used to make $w\sim w_0$, thus maximizing $\phi$. For a 3D lattice geometry, 
high N.A. is further needed to minimize phase shift contributions from non-target atoms. 
Different atoms could be measured by translating either the optical lattice
or the high finesse cavity, in two or three directions.
Assuming that the photons in the cavity are in a coherent state and that the detection 
of photons outside the cavity is shot-noise limited, the 
uncertainty of the phase shift $\Delta\phi$ depends on  
the number of photons in the cavity $N$ according to  
$\Delta \phi \propto \frac{1}{\sqrt{N}}$ \cite{Scully:97}.
This means that in order to successfully measure whether an atom is present in  
the cavity, we need to simultaneously satisfy two conditions.  
($i$) The phase shift has to be larger than its uncertainty, 
$\phi > \Delta\phi$. Using the equations above, this implies 
$N > \frac{\delta^2}{\Gamma^2 f^2}$. 
($ii$) The number of photons scattered from the measured atom has to be 
small, $N_{sc} < 1$. Using  
$N_{sc} = N(\frac{\Gamma}{\delta})^2f$, the condition ($i$) translates as
$N_{sc} > \frac{1}{f}$. For realistic cavities, $f$ can exceed $10^{5}$, so both conditions can be well satisfied.

A QND measurement can also be used to distinguish the magnetic field-sensitive leaked states from 
the qubit states, since they impose distinct phase shifts on a circularly polarized cavity field.
In the case of $^{87}Rb$, a qubit can leak to six other levels in the electronic ground state,
$m_{F=1} = \pm1$, $m_{F=2} = \pm1$ and $m_{F=2} = \pm2$. The states $m_{F=2} = \pm2$  cause opposite phase shifts of the same magnitude and hence can be individually resolved. 
The other states require a more subtle procedure because 
the levels $m_{F=1} = +1$ and $m_{F=1} = -1$ impose the same phase shift as 
$m_{F=2} = -1$ and $m_{F=2} = +1$ respectively. 
Assuming, for example, that the measurement  
outcome corresponds to the $m_{F=1} = +1$ and $m_{F=2} = -1$ states, one can first  
apply a state selective unitary operation \cite{Vala:03} which transforms the $m_{F=1} = +1$  
into the $m_{F=2} = 0$ state. 
A second QND measurement can now discriminate between 
these states. 
When a leaked state has been identified, i.e. the atomic wavefunction has collapsed 
to that particular quantum level, a short sequence of unitary operations can be made to rotate
the wavefunction from the leaked state into the qubit ground state. 
At this point, 
the situation is equivalent to the result of the first stage of qubit loss 
correction (see Eq.~(\ref{Eq:Sequence})). We note that an alternative approach to leakage error can be arrived at using a general leakage detection circuit 
with ancillary qubits as proposed in \cite{Preskill:97}. Application of 
this alternative approach to quantum computing with neutral atoms in an addressable optical lattice will be analyzed elsewhere \cite{Beals:IP}.

\paragraph*{Projective Measurement.} 
 
\begin{figure} 
\includegraphics[width=3in]{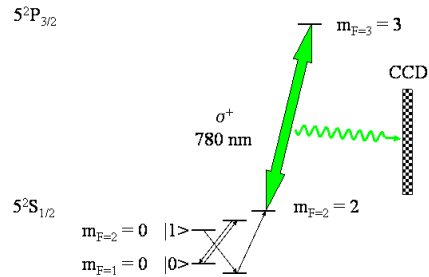} 
\caption{Scheme illustrating the projective measurement of a qubit encoded 
 in the hyperfine structure of the Rb electronic ground state manifold.} 
\label{Fig3} 
\end{figure}

Projective measurement can be realized using the cavity QED system. 
For example, the qubit levels which are stored originally in $m_F = 0$ states, 
can be distinguished by a QND measurement after unitary rotation into the
$m_{F=1} = 1$ and $m_{F=2} = 1$ levels, respectively. Since the cavity system is a part of the error correction setup, this solution may be more convenient than one based on fluorescence from the $5^2P_{3/2}$ state (illustrated in Fig. \ref{Fig3}).

Projective measurement is a rather demanding operation.  An alternative is 
a ``destructive'' measurement in which the particle carrying  
the qubit is ``destroyed'' during the measurement process, i.e. it is removed 
from the lattice after or before the measurement. 
There are several possible implementations of this option within the present scheme.  
The simplest is to make the ``destructive'' measurement, and then replace the atom.  The state can then be reset with a single qubit-flip operation conditional on the ``destructive'' measurement result. This procedure has the same effect in the error correction scheme as the original projective measurement.
 
\paragraph*{Conditional Source.} 
 
The conditional source can be implemented in an addressable optical lattice  
using an optical tweezer dipole trap to transport an atom in the qubit ground  
state $|0\ra$ from a reservoir lattice ('conveyor belt lattice') to  
the computation lattice. The dipole trap is loaded with an atom in the field  
insensitive state, $m_F = 0$, from an optical lattice made by the interference of two beams with perpendicular linear polarization vectors:  
$\theta = \pi /2$.  The dipole trap has a harmonic potential near the trap  
minimum and experiences translational displacement during tranportation  
of an atom but no change in well-depth or other parameters. If the  
timescale of the shift operation along one direction corresponds to an integer  multiple of the vibrational period of a trapped atom, then the extent of 
vibrational excitation during transportation can be effectively suppressed to  
zero \cite{Vala:03}.

\paragraph*{Conclusion}

We have presented a scheme for the correction of qubit loss and leakage errors during quantum computation with addressable atoms in optical lattices. It uses the GBP code to protect one logical qubit without any additional ancillary qubits and employs only a  small number of elementary quantum gates, together with a projective measurement and conditional source. It can also be applied in quantum information processing with trapped ions and photons.

Our research effort is sponsored by the Defense Advanced Research project Agency (DARPA) 
and the Air Force Laboratory, Air Force Material Command, USAF, under
contract No. F30602-01-2-0524.

\end{document}